\documentclass[preprint]{jpsj3}
\usepackage{txfonts}
\usepackage{amsmath}
\usepackage{bm}
\usepackage{braket}
\usepackage{color}
\usepackage{comment}

\newcommand\D{\mathrm{d}}
\newcommand\E{\mathrm{e}}

\usepackage{mathtools}
\DeclarePairedDelimiter{\abs}{\lvert}{\rvert}
\mathchardef\ELLSWAPTEMP=\mathcode`l
\def\ellon{%
\mathchardef\ell =\ELLSWAPTEMP
\mathcode`\l="0160
}

\def\elloff{%
\mathcode`\l=\ELLSWAPTEMP
\mathchardef\ell ="0160
}
\ellon

\title{Voltage-Controlled Magnonic Spin Tunneling Junction}

\author{Kohei Ohgane, Yuta Yahagi, Daisuke Miura\thanks{dmiura@solid.apph.tohoku.ac.jp} and Akimasa Sakuma}
\inst{Department of Applied Physics, Tohoku University, Sendai 980-8579, Japan}
\abst{
We theoretically investigate the effective exchange interaction, $J_\mathrm{eff}$, mediated by conductive electrons
within a nonmagnetic metal spacer, in the presence of a bias voltage, sandwiched by two ferromagnetic insulators.
On the basis of the tight-binding model,
we show the voltage and spacer thickness dependences of $J_\mathrm{eff}$,
and its contorollability is demonstrated.
We also propose a new magnonic device with the functions of both field effect transistor and non-volatile memory.
}

\begin{document}
\maketitle
\elloff

\begin{figure}[b]
\centering
\includegraphics[width=0.4\textwidth]{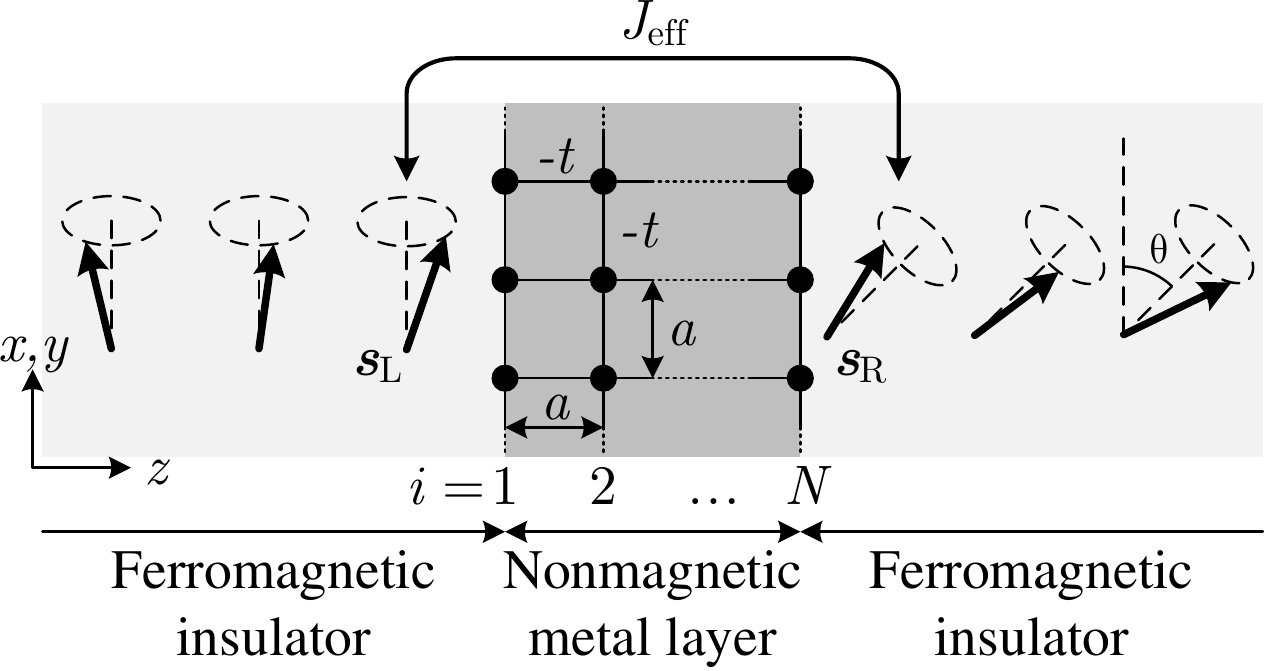}
\caption{Schematic view of the magnonic spin tunneling junction;
$\bm s_\mathrm{L(R)}$ denotes the direction of the localized spin on the left- (right-) hand-side interface,
$\theta$ is the relative angle between the magnetizations in the ferromagnetic insulators,
$J_\mathrm{eff}$ is the effective exchange interaction mediated by conductive electrons with the hopping integral $-t$,
and $a$ and $N$ represent the lattice constant and the thickness of the NM layer, respectively.
}
\label{fig:fig1}
\end{figure}

A magnonic spin tunneling junction (MSTJ) \cite{Ren2013} or a ferromagnetic insulating junction \cite{Nakata2018}
has the structure of FI/NM/FI where FI and NM represent a ferromagnetic insulator and a nonmagnetic spacer, respectively,
and functions as a spin Seebeck diode (SSD)\cite{Ren2013},
which has been expected as a fundamental element to realize integrated magnonic circuits.\cite{Kruglyak2010a,Chumak2014,Wang2018a}
The SSD effect is based on the spin Seebeck effect (SSE) in FIs,\cite{Uchida2010,Bauer2012,Adachi2013a}
and due to this phenomenon,
the amplitude of a tunneling magnon current driven by the SSE
passing through the NM spacer between the FIs depends on the direction of the thermal gradient applied on the MSTJ.
Ren and Zhu\cite{Ren2013} described the thermal-driven tunneling magnon current, $\bm I_\mathrm{s}$, in MSTJs and found the SSD effect.
Recently, we\cite{Miura2018b} extended their work to a case that magnetizations of FIs in MSTJs have a relative angle $\theta$, as shown in Fig. \ref{fig:fig1},
and demonstrated that a parallel condition ($\theta=0$) reproduces their result but an anti-parallel condition ($\theta=\pi$) gives $\bm I_\mathrm{s}=\bm 0$.
That is, the SSD element can be switched on/off by controlling the relative angle of magnetizations;
this tunable SSD effect was observed in YIG/Au/YIG\cite{Wu2018} and YIG/NiO/YIG\cite{Guo2018} junctions, at around the same time.

As a next step, we aim to control the SSD effect by external electric fields (voltage) toward electrically rewritable magnonic logic circuits.
Tang and Han investigated the tunneling magnon current passing through FI/FI/FI junctions
and suggested that the current could be controlled via a gate voltage induced Dzyaloshinskii--Moriya interaction\cite{Tang2019};
that is, the FI/FI/FI junction acts as a magnon field effect transistor (FET).
In this study, on the other hand, we assumed the use of a nonmagnetic metal as the NM spacer,
and attempted to control the SSD effect via a gate voltage modified Ruderman--Kittel--Kasuya--Yosida (RKKY) interaction.\cite{Ruderman1954,Kasuya1956,Yosida1957}
As shown in previous works\cite{Ren2013,Miura2018b},
the amplitude of $\bm I_\mathrm{s}$ is proportional to $\abs{J_\mathrm{eff}}^2$,
where $J_\mathrm{eff}$ denotes the strength of an effective exchange interaction between the FIs via the NM spacer.\cite{note1}
Then, assuming a voltaged NM spacer, we consider a possibility to control the SSD effect by using voltage-dependent $J_\mathrm{eff}$.
Although $J_\mathrm{eff}$ was given as a parameter in the previous works,
in order to reveal its voltage dependence,
we microscopically describe $J_\mathrm{eff}$ mediated by the conductive electrons in the voltaged NM spacer.


Let us consider the conductive electrons in the NM spacer described by the tight-binding Hamiltonian (see also Fig. \ref{fig:fig1}) as
\begin{align}
\mathcal{H}_\mathrm{NM}
&:=
\sum_{\bm k}\sum_{i=1}^{N}\epsilon_{\bm k} c_{\bm k i}^\dagger c_{\bm k i}
-t \sum_{\bm k}\sum_{i=1}^{N-1}
\left(
c_{\bm k i}^\dagger
c_{\bm k i+1}
+
c_{\bm k i+1}^\dagger
c_{\bm k i}
\right)
,
\end{align}
where
$c_{\bm k i}^\dagger (c_{\bm k i})$ is the creation (destruction) operator, in spinor representation,
of the electron with a wave vector $\bm k\equiv(k_x,k_y)$ in $i$th $x$--$y$ plane,
$N$ is the total number of NM layers,
$\epsilon_{\bm k}:=-2 t \cos k_x a-2 t\cos k_y a$ is the energy dispersion relation in the two dimensional square lattice of a lattice constant $a$,
and $-t$ is the hopping integral between the nearest sites.
We represent the effect of voltage by using a layer-dependent electric potential $\phi_i$,
that is, the potential energy due to voltage, $V$, is represented as
\begin{align}
\mathcal{V}:=
-e\sum_{\bm k}\sum_{i=1}^{N}\phi_i c_{\bm k i}^\dagger c_{\bm k i},
\end{align}
where $e>0$ is the elementary charge. We assume a simple form\cite{Ho1980,Fu1989,Neugebauer1993} of
\begin{align}
\phi_i=-\frac{V}{2}\frac{N-2 i+1}{N-1}.
\label{eq:phi}
\end{align}
Attaching the FIs on both sides ($i=1$ and $N$) of the above NM spacer,
the spins on the surfaces of the FIs interact with the conductive electrons on the surfaces of the NM spacer
via exchange interaction as\cite{Adachi2011,Ohnuma2017}
\begin{align}
\mathcal{H}_\mathrm{EX}:=
-J_\mathrm{L}\bm s_\mathrm{L}\cdot
\sum_{\bm k}c_{\bm k 1}^\dagger\hat{\bm\sigma} c_{\bm k 1}
-J_\mathrm{R}\bm s_\mathrm{R}\cdot
\sum_{\bm k}
c_{\bm k {N}}^\dagger\hat{\bm\sigma} c_{\bm k {N}},
\end{align}
where
$\hat{\bm \sigma}$ is the Pauli matrices,
$\bm s_\mathrm{L (R)}$ is the unit vector of direction of spin
on the left- (right-) hand-side of the FI interface, in which the strength of interaction is denoted by $J_\mathrm{L(R)}$.

\begin{figure}[tb]
\centering
\includegraphics[width=0.45\textwidth]{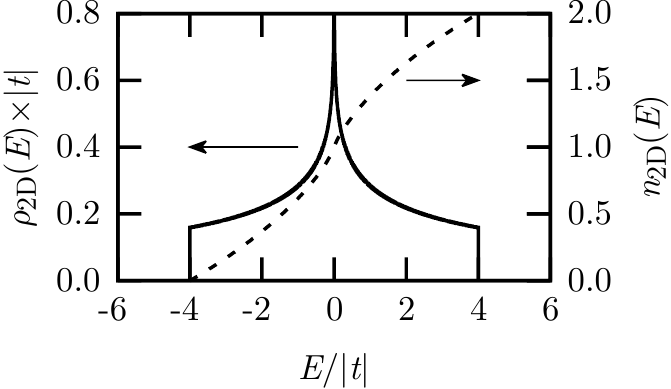}
\caption{The solid and dashed lines are the density of states $\rho_\mathrm{2D}(E)$ and the electron density $n_\mathrm{2D}(E)$, respectively, in the two-dimensional square lattice.}
\label{fig:fig2}
\end{figure}

In order to obtain the voltage-dependent $J_\mathrm{eff}$,
we performed perturbative expansion with respect to $\mathcal{H}_\mathrm{EX}$ for the Helmholtz free energy of the system.
Because the first-order perturbation energy vanishes,
the lowest contribution appears in second-order $\mathcal{H}_\mathrm{EX}$ and is represented as
\begin{align}
\varDelta F=-\frac{1}{2}\int_0^\beta\D\tau
\Braket{\E^{\tau (\mathcal{H}_\mathrm{NM}+\mathcal{V})}
\mathcal{H}_\mathrm{EX}
\E^{-\tau (\mathcal{H}_\mathrm{NM}+\mathcal{V})}
\mathcal{H}_\mathrm{EX}
},
\end{align}
where $\beta$ denotes the inverse temperature,
$\braket{\cdots}:=\Tr\E^{-\beta(\mathcal{H}_\mathrm{NM}+\mathcal{V}-\mu)}\cdots/\Tr\E^{-\beta(\mathcal{H}_\mathrm{NM}+\mathcal{V}-\mu)}$, and the chemical potential $\mu$ is determined for a fixed electron density.
Taking the low-temperature limit ($\beta\to\infty$) after the thermodynamic limit ($M:=\sum_{\bm k}1\to\infty$),
we obtain
\begin{align}
\lim_{\beta\to\infty}\lim_{M\to\infty}\frac{\varDelta F}{M}
=
u_0-J_\mathrm{eff}\bm s_\mathrm{L}\cdot\bm s_\mathrm{R},
\end{align}
where $u_0$ is an independent part of $\bm s_\mathrm{L}$ and $\bm s_\mathrm{R}$,
and we defined the effective exchange coupling constant via the NM spacer as
\begin{align}
J_\mathrm{eff}&:=J_\mathrm{L} J_\mathrm{R}\sum_{p}
y_1^p
y_{N}^{p}
\Biggl(
y_1^{p}
y_{N}^{p}
\rho_\mathrm{2D}(\mu-\lambda_p)
\notag\\
&-
\sum_{p'\neq p}
y_1^{p'}
y_{N}^{p'}
\frac{n_\mathrm{2D}(\mu-\lambda_p)-n_\mathrm{2D}(\mu-\lambda_{p'})}{\lambda_p-\lambda_{p'}}
\Biggr),
\label{eq:jeff}
\end{align}
For simplicity, we use the normalized one defined as
\begin{align}
\tilde J_\mathrm{eff}:=\frac{\abs{t}}{J_\mathrm{L}J_\mathrm{R}} J_\mathrm{eff}.
\label{eq:tjeff}
\end{align}
In Eq. (\ref{eq:jeff}), $\lambda_p$ and $(y_1^p,y_2^p,\ldots,y_{N}^p)$, respectively, are the eigenvalue and its corresponding eigenvector in the equation,
\begin{align}
(-e\phi_i-\lambda_p) y_{i}^p
-t y_{i-1}^p-ty_{i+1}^p=0
\quad\mathrm{with}\quad
y_0^p=y_{N+1}^p=0,
\label{eq:eigeneq}
\end{align}
which is derived from the Schr\"odinger equation for the Hamiltonian $\mathcal{H}_\mathrm{NM}+\mathcal{V}$
and has the information on the bias voltage and the connection between layers.
$\rho_\mathrm{2D}(E)$
and
$n_\mathrm{2D}(E)$,
respectively,
are the density of states (DOS) at $E$ and the number of the electrons occupying the states below $E$, per site in a two-dimensional square lattice.
They are defined by
\begin{align}
\rho_\mathrm{2D}(E):=\lim_{M\to\infty}\frac{2}{M}\sum_{\bm k}\delta(E-\epsilon_{\bm k})
\end{align}
and
\begin{align}
n_\mathrm{2D}(E):=\int_{-\infty}^E \D E' \rho_\mathrm{2D}(E'),
\end{align}
Furthermore, $\rho_\mathrm{2D}(E)$ can be represented as\cite{Pesz1986}
\begin{align}
\rho_\mathrm{2D}(E)=\frac{1}{\pi^2\abs{t}}K\left(\sqrt{1-\frac{E^2}{16t^2}}\right)\theta(4\abs{t}-\abs{E}),
\end{align}
in terms of the elliptic integral of the first kind:
\begin{align}
K(k):=\int_0^1\frac{\D x}{\sqrt{(1-x^2)(1-k^2 x^2)}}.
\end{align}
Their energy dependences are shown in Fig. \ref{fig:fig2}.

Before discussing the voltage dependence of the effective exchange interaction,
we focus on inversion symmetry of the electric potential, $\phi_i=-\phi_{N+1-i}$, as represented by Eq. (\ref{eq:phi}).
It leads to inversion symmetry with respect to $V$,
\begin{align}
\tilde J_\mathrm{eff}\bigr|_{V}=\tilde J_\mathrm{eff}\bigr|_{-V},
\end{align}
and the electron--hole symmetry,
\begin{align}
\tilde J_\mathrm{eff}\bigr|_{\mu}=\tilde J_\mathrm{eff}\bigr|_{-\mu}.
\end{align}
Therefore, we only consider $\mu \le 0$ and $V\ge 0$.

Firstly, we consider the zero bias case of $V=0$ (i.e., $\phi_i=0$).
In this case, Eq. (\ref{eq:eigeneq}) can be solved exactly (for example, p. 137 in Ref. \citenum{Gregory1978}),
and its solution is expressed as
\begin{subequations}
\begin{align}
\lambda_p&\to \lambda_p^0:=-2\abs{t}\cos\left(\frac{p\pi}{N+1}\right),\\
y_i^p&\to\Biggl(\frac{t}{\abs{t}}\Biggr)^i\sqrt{\frac{2}{N+1}}\sin\left(\frac{i p \pi}{N+1}\right),
\end{align}
\label{eq:exact}
\end{subequations}%
where $p=1,2,\ldots,N$.
Applying Eq. (\ref{eq:exact}) to Eq. (\ref{eq:tjeff}), and then, letting the result be $\tilde J_\mathrm{eff}^0$ and Fermi level be $\mu^0$,
we have
\begin{align}
&\tilde J_\mathrm{eff}^0
=
\frac{4\abs{t}}{(N+1)^2}
\sum_{p=1}^N
\Biggl(
\sin^4\frac{p\pi}{N+1}
\rho_\mathrm{2D}(\mu^0-\lambda_p^0)
\notag\\
&-\frac{3}{2\abs{t}}
\sin^2\frac{p\pi}{N+1}
\cos\frac{p\pi}{N+1}
n_\mathrm{2D}(\mu^0-\lambda_p^0)
\Biggr),
\end{align}
which corresponds to the RKKY interaction in terms of the tight-binding scheme.
We notice that the half-filling condition $\mu^0=0$ yields $\tilde J_\mathrm{eff}^0\to +\infty$ for any odd number $N$ due to $\rho_\mathrm{2D}(0)\to+\infty$
because $p$ exists as $\lambda_p^0=0$.
In this condition, we cannot justify the perturbative treatment with respect to $H_\mathrm{EX}$ (and cannot obtain any finite result for $V$).
However, for any even number $N$, we can get a finite $\tilde J_\mathrm{eff}^0$, as expected.

Next, we consider the effect of voltage on the effective exchange interaction.
The normalized voltage is defined as
\begin{align}
v:=\frac{eV}{\abs{t}},
\end{align}
and, when evaluating $\tilde J_\mathrm{eff}$, we use voltage-dependent $\mu$ satisfying
\begin{align}
\frac{1}{N}\sum_p n_\mathrm{2D}(\mu-\lambda_p)
=\frac{1}{N}\sum_p n_\mathrm{2D}(\mu^0-\lambda_p^0)
\equiv n^0
,
\label{eq:mu}
\end{align}
because the electron density $n^0$ given by $\mu^0$ must be invariant for $v$.
\begin{figure}[tb]
\centering
\includegraphics[width=0.38\textwidth]{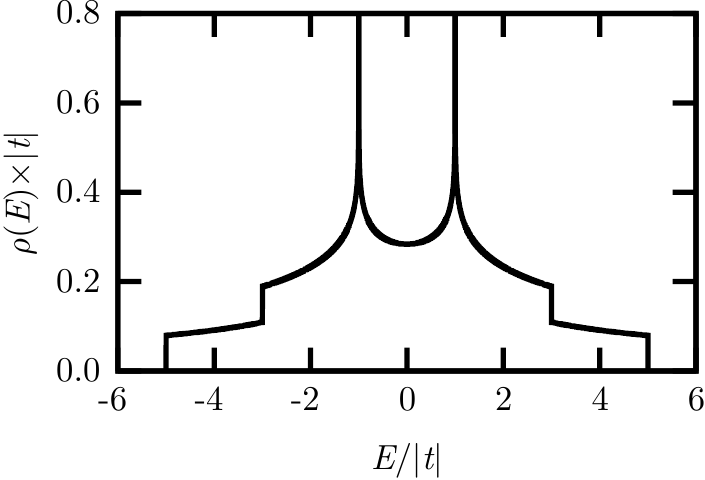}
\caption{
The total density of states $\rho(E)$ as a function of $E$ in a two-layered system in zero bias.
}
\label{fig:fig3}
\end{figure}

The two-layer case ($N=2$) is not only the simplest and but also demonstrates our main results.
Solving Eq. (\ref{eq:eigeneq}) with $\phi_1=-V/2$ and $\phi_2=V/2$,
we get $\lambda_1=-\lambda_2\equiv -\lambda$
and $y_1(1)^2y_2(1)^2=y_1(2)^2y_2(2)^2=-y_1(1)y_1(2)y_2(1)y_2(2)=-y_1(2)y_1(1)y_2(2)y_2(1)=t^2/(4\lambda^2)$,
where $\lambda:=\abs{t}\sqrt{1+v^2/4}$.
Therefore, we have
\begin{align}
\tilde J_\mathrm{eff}
=
\frac{\abs{t}^3}{2 \lambda^2}
\left[
\rho(\mu)
-
\bar\rho(\mu)
\right],
\label{eq:jeff2}
\end{align}
where,
\begin{align}
\rho(E)&:=\frac{\rho_\mathrm{2D}(E+\lambda)+\rho_\mathrm{2D}(E-\lambda)}{2},\label{eq:rho}\\
\bar\rho(E)&:=\frac{1}{2\lambda}\int_{E-\lambda}^{E+\lambda}\D E' \rho_\mathrm{2D}(E'),\label{eq:barrho}
\end{align}
Furthermore, the condition (\ref{eq:mu}) for $\mu$ should also be considered.
From the definitions (\ref{eq:rho}) and (\ref{eq:barrho}),
we can see that
$\rho(\mu)$ means the arithmetic average of $\rho_\mathrm{2D}(E)$ at $E=\mu+\lambda$ and $E=\mu-\lambda$ (or, the total DOS at the Fermi level)
and
$\bar\rho(\mu)$ means the average of $\rho_\mathrm{2D}(E)$ over the interval $[\mu-\lambda,\mu+\lambda]$.
Figure \ref{fig:fig3} shows $\rho(E)$ as a function of $E$,
in which
we can observe that $\rho(E)$ has four discontinuities and two spikes caused by the singularities of $\rho_\mathrm{2D}(E)$, as shown in Fig. \ref{fig:fig2} (solid line).
Therefore, the $\rho(\mu)$ term in Eq. (\ref{eq:jeff2}) is also discontinuous at the corresponding $\mu$.
On the other hand, $\bar\rho(\mu)$ is continuous for $\mu$.
On the basis of the above features,
let us discuss the voltage dependence of $\tilde J_\mathrm{eff}$ for several values of $\mu^0$.
\begin{figure}[tb]
\centering
\includegraphics[width=0.35\textwidth]{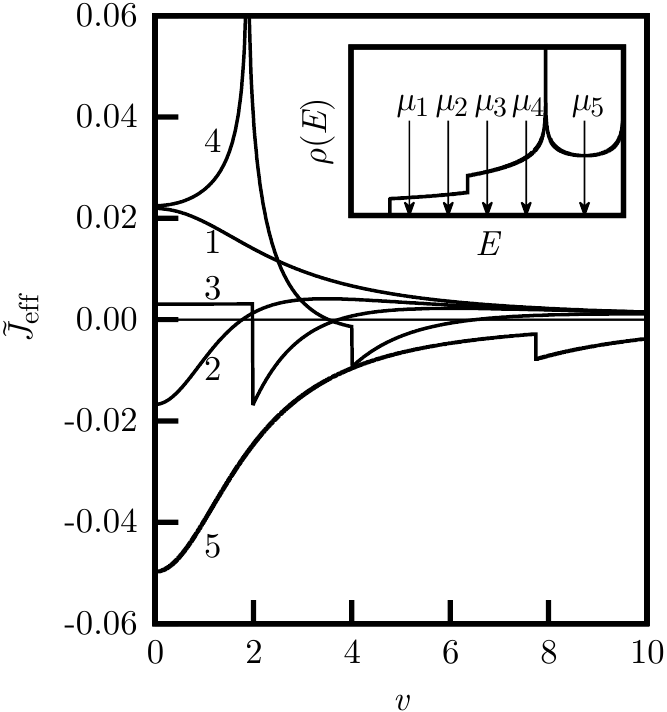}
\caption{The calculated normalized effective exchange interaction $\tilde J_\mathrm{eff}$ as a function of normalized voltage $v$ for several $\mu^0=\mu_i$
in the two-layer system ($N=2$), where $i$ is denoted by the number of each line.
The $\mu_i$'s are given as
$\mu_1=-4.5$,
$\mu_2=-3.5$,
$\mu_3=-2.5$,
$\mu_4=-1.5$,
and $\mu_5=   0$
in units of $\abs{t}$.
The inset shows the position of $\mu_i$ in the energy dependence of the total DOS $\rho(E)$.
}
\label{fig:fig4}
\end{figure}
Here, we examine $\mu^0/\abs{t}\ (\mathrm{or}\ n^0)=-4.5\ (0.0411), -3.5\ (0.133), -2.5\ (0.284), -1.5\ (0.609)$, and $0 (1)$,
whose positions within the band width in the system are shown by $\mu_1, \mu_2, \mu_3, \mu_4,$ and $\mu_5$, respectively, in the inset in Fig. \ref{fig:fig4}.
Lines 1 and 2 in the main figure in Fig. \ref{fig:fig4}
are on the first shoulder of $\rho(E)$, originating from the lower band $\epsilon_{\bm k}+\lambda$,
in which $\mu$ does not touch the second shoulder of $\rho(E)$ for any $v$.
Therefore, $\tilde J_\mathrm{eff}$ is continuous for $v$.
In addition, we observe that $\tilde J_\mathrm{eff}$ becomes positive for a large $v$, indicating
that $\tilde J_\mathrm{eff}\to 2 \abs{t}\rho(\mu)/v^2>0 $ for $v\to \infty$.
That is, in the condition such that $\rho(\mu^0)-\bar\rho(\mu^0)<0$,
the sign of $\tilde J_\mathrm{eff}$ changes from negative to positive,
and a voltage exists such that $J_\mathrm{eff}=0$; except for the half-filling case.
Lines 3 and 4 in the main figure in Fig. \ref{fig:fig4} are examples that $\mu^0$ is on the second shoulder of $\rho(E)$.
In line 3,
$\mu$ drops from the second shoulder to the first one of $\rho(E)$ at $v\simeq 2$,
and as a result, $\tilde J_\mathrm{eff}$ drastically drops at $v\simeq 2$ and tends to lines 1 and 2 with increasing $v$.
In line 4, the spike of $J_\mathrm{eff}$ at $v\simeq 2$ originates from that $\mu$ touches the spike of $\rho(E)$,
and the drop at $v\simeq 4$ originates from that $\mu$ moves from the second shoulder to the first one of $\rho(E)$;
thus, it also tends to lines 1 and 2.
Line 5 is in the half-filling state $n^0=1$ (or $\mu^0=0$),
in which $\mu$ is independent of $v$ (i.e., $\mu=\mu^0=0$) and Eq. (\ref{eq:jeff2}) is simplified to
\begin{align}
\tilde J_\mathrm{eff}
=
\frac{\abs{t}^3}{2\lambda^2}
\left(
\rho_\mathrm{2D}(\lambda)
-
\frac{1}{\lambda}
\int_0^\lambda
\D E
\rho_\mathrm{2D}(E)
\right).
\end{align}
From this expression,
it is implied
that $J_\mathrm{eff}<0$ for any $v$ because $\rho_\mathrm{2D}(E)$ for $E\ge 0$ monotonically decreases with increasing $E$,
and that $J_\mathrm{eff}$ is discontinuous at $v=2\sqrt{15}$ where $\mu$ enters into the band gap.
As shown above,
we can control $J_\mathrm{eff}$ well by the voltage in the sense that its sign is controllable,
or if desired, we can also protect one from the variance of the voltage,
by tuning $\mu^0$.

\begin{figure}[tb]
\centering
\includegraphics[width=0.35\textwidth]{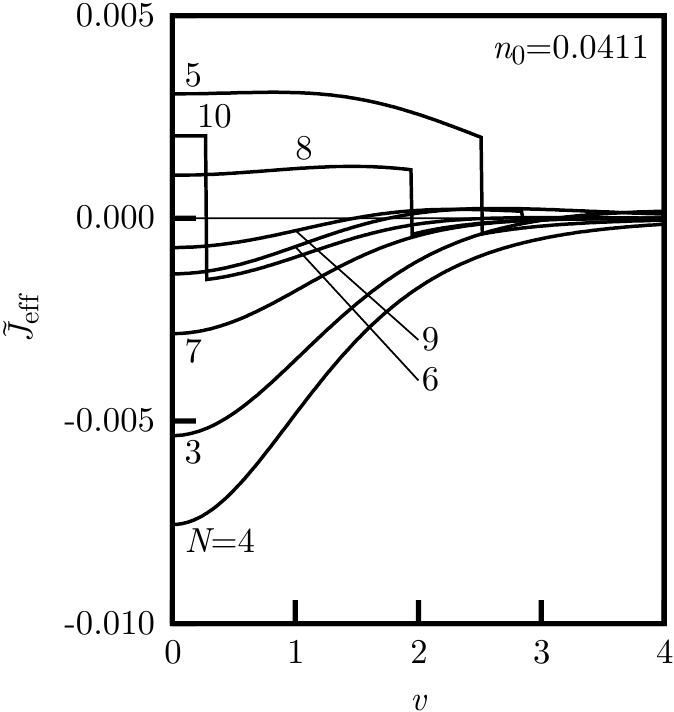}
\caption{The calculated normalized effective exchange interaction $\tilde J_\mathrm{eff}$ as functions of normalized voltage $v$
in $N$-layer systems.
In the calculations, the electron density is fixed to be $n_0=0.0411$,
which corresponds to $\mu^0/\abs{t}=\mu_1/\abs{t}=-4.5$ in the $N=2$-layer system
(see line 1 in the main figure in Fig. \ref{fig:fig4}).
}
\label{fig:fig5}
\end{figure}
\begin{figure}[tb]
\centering
\includegraphics[width=0.4\textwidth]{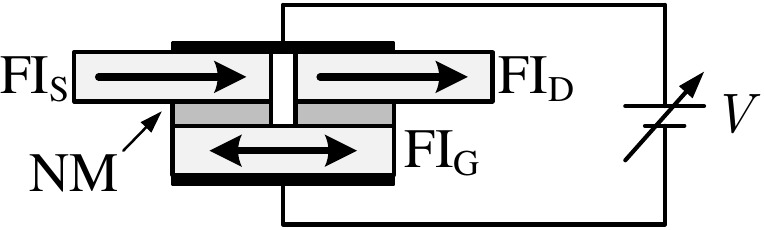}
\caption{Schematic view of new magnonic device with the functions of both FET and non-volatile memory.
}
\label{fig:fig6}
\end{figure}
%
%


Lastly, we explore the dependence of the exchange effective interaction on the NM spacer thickness $N$.
In general, $\tilde J_\mathrm{eff}$ tends to be small with increasing $N$
because $\tilde J_\mathrm{eff}$ is dominated by the product of wave functions $y_1^p y_N^{p'}$,
which decays with increasing distance between the sites $1$ and $N$, as shown in Eq. (\ref{eq:jeff}).
However, the electronic structure also depends on $N$, for example, the total DOS in an $N$-layer system
has $N$ spikes in energy dependence.
Thus, the voltage dependence of $\tilde J_\mathrm{eff}$ strongly depends on $N$ for a fixed electron density $n_0$.
Figure \ref{fig:fig5} shows the voltage dependence of $\tilde J_\mathrm{eff}$ for $N=3,4,\ldots,10$ in the fixed electron density $n_0=0.0411$.
We can observe that
the various behaviors in $\tilde J_\mathrm{eff}$ vs. $v$ are available by increasing $N$
although the result for $N=2$ is always positive and monotonically decreases with increasing $v$ (see the line 1 in Fig. \ref{fig:fig4}).
These results suggest that one can handle the voltage controllability of $\tilde J_\mathrm{eff}$ by tuning the NM spacer thickness.

As an example applying the above results,
we propose the new magnonic device shown in Fig. \ref{fig:fig6},
which provides both FET and non-volatile memory functions.
(1) magnon FET: when the gate voltage $V\neq 0$ and the magnetization direction of the FI$_\mathrm{G}$ is parallel to ones of the FI$_\mathrm{S}$ and FI$_\mathrm{D}$,
the magnon current from the FI$_\mathrm{S}$ to FI$_\mathrm{D}$ is controlled by the gate voltage via the RKKY interaction.
(2) non-volatile magnon memory:
when $V=0$ and the FI$_\mathrm{G}$ is in the parallel (or antiparallel) condition,
the magnon current is permitted (or blocked) by the tunable SSD effect\cite{Wu2018,Guo2018,Miura2018b};
that is, the FI$_\mathrm{G}$ can hold 1 bit of data.

In summary, we described the effective exchange interaction via a voltaged NM spacer within the tight-binding model
to consider a possibility to control it electrically.
As a result, we showed that the voltage dependence of the effective exchange interaction strongly depends on the chemical potential and NM spacer thickness,
and it was suggested that one can control the effective exchange interaction well through the RKKY interaction modified by voltage.
Furthermore, we proposed a new magnonic device providing both FET and non-volatile memory functions.

\begin{acknowledgment}
This work was supported by JSPS KAKENHI Grant No. 17K14800 in Japan, and Center for Spintronics Research Network (CSRN).
\end{acknowledgment}

\bibliographystyle{jpsj}
\bibliography{library,FOOTNOTES}

\end{document}